\documentclass[aps,prl,twocolumn,superscriptaddress]{revtex4-1}
\usepackage{amssymb}
\usepackage{amsmath}
\usepackage{graphicx}% Include figure files
\usepackage{dcolumn}% Align table columns on decimal point
\usepackage{bm}% bold math
\usepackage{dsfont}
\usepackage{subcaption}

\usepackage[utf8]{inputenc}
\usepackage[T1]{fontenc}
\usepackage{newtxtext,newtxmath}

\usepackage[dvipsnames]{xcolor}

\begin{document}
\title{
% The Redundancy principle and the role of heterogeneity - PRL
% \\
% \dhc{Heterogeneity in redundancy, one winner takes it all}
% Population heterogeneity and the redundancy principle
The impact of population heterogeneity on the redundancy principle
}

\author{Sergei Fedotov}
\email[]{sergei.fedotov@manchester.ac.uk}
\affiliation{Department of Mathematics, University of Manchester, M13 9PL, UK}

\author{Daniel Fears}
%\email[]{sergei.fedotov@manchester.ac.uk}
\affiliation{Department of Mathematics, University of Manchester, M13 9PL, UK}

\author{Daniel Han}
% \email[]{}
\affiliation{School of Mathematics \& Statistics,
UNSW, Sydney, Australia}

%\date{}

\begin{abstract}
    Biological signaling is often governed by extreme value statistics, where a rapid response relies on the fastest few out of a large redundant group of searchers.
    While extreme first passage time (FPT) theory is well established for homogeneous ensembles, its sensitivity to population heterogeneity remains open.
    We show that averaging over a heterogeneous population of memoryless random walkers gives rise to ensemble self-reinforcement.
    This heterogeneity drastically changes both the FPT and minimum FPT densities relative to a homogeneous ensemble with identical mean rates.
    The modal and minimum FPTs are an order of magnitude smaller for heterogeneous populations relative to homogeneous ones.
    Our exact analytical predictions establish that population heterogeneity is a parameter that biology can exploit and not merely noise to be averaged away.

\end{abstract}

\maketitle

Across scales, critical processes in biology are governed by the fastest few. 
One spermatozoon out of $\sim 10^7$ fertilizes the egg  \cite{reynaud2015so}; a handful of motor proteins out of $\sim 10^6$ suffice to move an intracellular vesicle along its microtubule track \cite{hollenbeck1989distribution,klumpp2005cooperative,reck2018cytoplasmic}; a few calcium ions out of $\sim 10^3$ trigger the avalanche of calcium-induced calcium release in a dendritic spine of a neuron \cite{higley2012calcium,basnayake2019fast}.
For each process, random walks are performed to search for the target governed by thermodynamic principles, and almost all will fail.
However, this redundancy is nature's strategy for generating a reliable and timely response to critical biological processes \cite{meerson2015mortality,schuss2019redundancy}.
These observations imply that some timescales of life are regulated by the extremes of the first passage time (FPT) distribution \cite{resnick1987extreme,schuss2019redundancy} rather than their means, which remain important in other contexts \cite{holcman2014narrow,metzler2014first,guerin2016mean}.

Properties of the minimum FPT have been well studied in a variety of domains \cite{basnayake2019asymptotic} and a universal asymptotic formula exists \cite{lawley2020universal}. 
The standard method to obtain analytical results \cite{basnayake2019asymptotic,lawley2020universal} is to assume $N$ independent and identically distributed (i.i.d.) random walkers that have FPTs, $t_1,\cdots,t_N$ to find a target. 
Then one defines the minimum FPT out of $N$ random walkers to be $\tau_N = \min\left\{t_1,\cdots,t_N\right\}$.
Recent theoretical work has shown that the minimum FPT of Brownian random walkers to reach a target in one dimension scales as $\tau_N\sim L^2/\left[4D \ln(N)\right],$
where $L$ is the length of a domain and $D$ is the diffusion coefficient \cite{lawley2020universal} with similar results in two and three spatial dimensions \cite{basnayake2019asymptotic}.

However, these minimum FPT results rest upon the assumption that each stochastic realization of the random walk is governed by the same rules.
While the assumption is mathematically convenient, the heterogeneous nature of biological processes necessitates the relaxation of this assumption.
The minimum FPTs scale with stronger dependence on particle numbers when the initial condition is uniformly distributed on a domain as opposed to all random walks starting at a single position \cite{grebenkov2020single}.
This sensitive dependence on initial conditions suggests that heterogeneity is critical when considering minimum FPTs.
Just as the mean FPT misrepresents the time scales of redundant searches, the homogeneous population assumption collapses minimum FPT statistics.

There is substantial experimental evidence to suggest that biology favors population heterogeneity.
Sperm populations contain distinct motility-type subgroups, where only a minority subpopulation is consistently superior for fertilization \cite{holt2004concepts,fernandez2022predicting}.
Intracellular vesicles exhibit broad multi-modal distributions when diffusion coefficients of single particle trajectories are measured with cargo variability dominating population-level kinetics \cite{wang2012brownian,han2020deciphering,sabri2020elucidating,korabel2021local}.
Dendritic spine calcium transients show that variability in stochastic dynamics and spine morphology affects spiking and plasticity in the brain \cite{higley2012calcium,anwar2013stochastic,basnayake2019fast}.
Just as the minimum FPT is more informative than the mean FPT, the biological feature of interest is not the mean behavior of a homogeneous ensemble but the distribution of behaviors in a heterogeneous population.
From spermatozoon to calcium ions, experimental evidence across biological scales converges to the fact that the searcher who finds the target first is a consistent winner out of a heterogeneous population designed to operate above biological noise levels.
In fact, it has been shown in the context of diffusing diffusivity that heterogeneity allows non-Gaussian diffusion to have minimum FPTs shorter than Gaussian diffusion by orders of magnitude \cite{sposini2024being}.
To understand this phenomenon and how the minimum FPT results change when heterogeneous random walkers are introduced, we require analyses on the mesoscopic level and not in the diffusion limit.

In this paper, we consider a heterogeneous ensemble of random walks at the mesoscopic scale whose rate of jumping and spatial bias are random variables.
By formulating the master equation for this heterogeneous ensemble and solving it analytically, we show that the averaged probability mass function exhibits ensemble self-reinforcement.
Furthermore, we show that introducing heterogeneity into a random-walker ensemble reduces  both the first-passage time and its minimum by an order of magnitude relative to a homogeneous ensemble with identical average jump rates and biases.

We start by examining the master equation for the probability
\(
p(k,t)=\mathbb{P}\{ X(t)=k \},
\)
where $X(t)$ is the position of a particle at point $k\in \mathbb{Z}$  at time $t$. 
The master equation is
\begin{equation} \label{masterequation random rates}
\frac{\partial p}{\partial t}=-(\gamma_+   +\gamma_- )p(k,t)
+ \gamma_+ p(k-1,t) +\gamma_-   p(k+1,t)
\end{equation}
where $\gamma_+$ and $\gamma_-$ are the random rate of jumps to the right (+) and left (-), respectively. 
We assume that $\gamma_+$ and $\gamma_-$ are independent gamma distributed with probability density function 
\begin{equation}
    f(\gamma_\pm) = \frac{\beta^{\alpha_\pm}\gamma_\pm ^{\alpha_\pm - 1}e^{-\beta\gamma_\pm}} {\Gamma(\alpha_\pm)}  \hspace{.2in}
\gamma_\pm \geq 0.
\label{eq:GammaPDF}
\end{equation}
For a biological context, Eq. \eqref{masterequation random rates} can model intracellular vesicle transport along microtubules with $\gamma_{\pm}$ representing the effective stepping rates arising from the multiple kinesin and dynein motor proteins bound to each cargo \cite{klumpp2005cooperative,muller2008tug,berger2012distinct}. 
As the effective stepping rate represents a combination of biological contributions such as motor protein number, adaptor binding, dynein positioning and ATP availability, the variability across the population of intracellular vesicles is naturally characterized by the gamma distribution \eqref{eq:GammaPDF}.
This assumption is consistent with experimental observations of gamma distributed velocities obtained through single molecule tracking in live cell vesicle transport microscopy \cite{shen2025stick}.
The independence of $\gamma_+$ and $\gamma_-$ is well motivated as kinesin and dynein motors are recruited via distinct biochemical pathways \cite{vale2003molecular}.

It is convenient to rewrite the master equation as
\begin{equation}
\frac{\partial p}{\partial t}=-\gamma p(k,t)+ \gamma q
 p(k-1,t) +\gamma (1-q)  p(k+1,t).
 \label{eq:continuousmaster}
\end{equation}
Here, \(\gamma=\gamma_++\gamma_-\) represents the total jump rate, and \(q=\gamma_+/\gamma\) is the right-ward jump probability. By rewriting the master equation in such a form, we can use  Lukacs's proportion-sum independence theorem \cite{lukacs1955characterization,mosimann1962compound} to consider \(\gamma\) and \(q\) as independent random variables following a gamma and a beta distribution, respectively:
\begin{equation}
    \xi(q) = \frac{q^{\alpha_+ - 1}(1 - q)^{\alpha_- - 1}}{B(\alpha_+,\alpha_-)}, \quad \zeta(\gamma)=\frac{\beta^{\alpha}\gamma^{\alpha- 1}e^{-\beta\gamma}}{\Gamma(\alpha)},
\end{equation}
where \(\alpha=\alpha_+ + \alpha_-\).

The solution to \eqref{eq:continuousmaster} can be written in terms of two underlying random processes: a discrete random walk, with position \(X_n\), and a Poisson jump process, \(N(t)\), via subordination. We set \(X(t) = X_{N(t)}\), giving the solution to the master equation \eqref{eq:continuousmaster}, \(p(k,t \mid q, \gamma)=\sum_{n=0}^{\infty}P(k,n \mid q)  (\gamma t) ^ {n} e ^ {-\gamma t}/n!\), where \(P(k,n \mid q)\) is the well-known solution of the discrete walk conditioned on \(q\) \cite{feller1968introduction}, \(P(k,n \mid q) = \binom{n}{\frac{1}{2}(n+k)} q^{\frac{1}{2}(n+k)}(1-q)^{\frac{1}{2}(n-k)}\). Averaging this conditional probability mass function, we obtain  
\begin{equation} \label{eq: pbar averaged}
    \bar{p}(k,t) = \mathbb{E}_{q,\gamma} [p(k,t \mid q, \gamma)] = \sum_{n=0}^\infty \bar{P}(k,n)\,Q(n,t),
\end{equation}
where
\begin{equation} \label{eq: p bar discrete walk average}
    \bar{P}(k,n) = \binom{n}{\frac{1}{2}(n+k)}\frac{B\left(\alpha_++\frac{1}{2}(n+k), \alpha_-+\frac{1}{2}(n-k)\right)}{B(\alpha_+,\alpha_-)}
\end{equation}
and 
\begin{equation} \label{eq: Q equation}
    Q(n,t) = \frac{\beta^\alpha t^n}{n!}\frac{\Gamma(\alpha+n)}{\Gamma(\alpha)}(\beta+t)^{-(\alpha+n)}. 
\end{equation}

This averaging marks a drastic shift in the behavior of the system: the underlying random processes---individually memoryless discrete random walks driven by a P\'olya counting process---together exhibit self-reinforcement, with dramatic consequences for first passage times and extreme statistics. 
The underlying discrete random walk described by $\bar{P}(k,n)= \mathbb{E}_{q,\gamma}[P(k,n \mid q)] = \int_{0}^{1} P(k,n \mid q) \xi(q) dq$
follows a discrete master equation with self-reinforcement: \(\bar{P}(k,n+1)=u_{n}^{+}(k-1)
 \bar{P}(k-1,n) +u_{n}^{-}(k+1)\bar{P}(k+1,n)  \)
where the transition probabilities are \(u_{n}^{\pm}(k)=\frac{1}{2}\pm\frac{1}{2} g(k,n)\) where \(g(k,n) =  (\alpha_+ -\alpha_-+k)/(\alpha_+ +\alpha_- +n)    \) \cite{PhysRevE.107.034115} (see Appendix).
Self-reinforcement in this context means, for example, the probability of jumping on the right, \(u_n^+\), is an increasing function in \(k\), so a particle that has drifted to the right has, on average, a higher effective rightward bias, and is therefore more likely to continue rightward. 
Heterogeneity also leads to self-reinforcement in the counting process: instead of a Poisson process, it is now described by a P\'olya process. The function \(Q(n,t) = \mathbb{P}(N(t) = n)\) defined in \eqref{eq: Q equation} obeys the Kolmogorov forward equation (master equation) for the P\'olya process $N(t)$ \(d Q(n,t)/dt = \lambda_{n-1}(t) \,Q(n-1,t)
- \lambda_n(t) \,Q(n,t)\)
with rate \(\lambda_n(t) = (\alpha +n)/( \beta + t)\) (see Appendix).
This rate increases with the number of jumps, \(n\), so each jump raises the likelihood of the next, generating temporal self-reinforcement, complementary to the spatial self-reinforcement described above. 
As a consequence of this ensemble self-reinforcement, the ensemble averaged position of the particle, \(\bar{X}(t)\), exhibits ballistic superdiffusion, characterized by the quadratic variance \(\mathrm{Var}(\bar{X}(t)) = \alpha \, t^2/\beta^2 + \alpha \, t/\beta\) (see Appendix). This behavior is a consequence of parameter heterogeneity, rather than environmental factors, and is consistent with experimental observations of intracellular transport \cite{Gavrilova2025}.
Self-reinforcement fundamentally changes first-passage and extreme statistics compared to homogeneous random walks.

Now, we aim to see the impact of heterogeneity on first passage times for the particles. First, we consider a particle with fixed values of \(q\) and \(\gamma\), starting from the position \(X(0)=0\) on a semi-infinite lattice, ranging from \(-\infty\) to a positive boundary \(k=m\).
The continuous-time for a particle to first reach the boundary at \(m\) is \(T_m = \inf\{t \ge 0: X(t) = m\}\). It is convenient to introduce the first passage time for the discrete random walk, \(X_n\). For fixed \(q\), we define the discrete first passage time to level \(m\) as \(\tau_m = \inf\{n \ge 0: X_n = m\}\). Thus, \(T_m\) is defined by the sum \(T_m = \sum_{k=1}^{\tau_m} J_k\), where \(J_k\) are the i.i.d. waiting times between jumps, with common density \(\psi(t)\), independent of \(\tau_m\). By the law of total probability, the FPT density for fixed \(q\) and \(\gamma\) is
\begin{equation}
    f_{m}(t \mid q, \gamma) = \sum_{n=m}^{\infty} \mathbb{P}\{\tau_m = n \mid q\} \, \psi^{*n}(t) ,
    \label{eq: FPT general formula}
\end{equation}
where \(\psi^{*n}\) denotes the \(n\)-fold convolution of \(\psi\) and the sum runs over \(n \ge m\) with \(n+m\) even. As the waiting times between jumps are exponentially distributed with rate \(\gamma\), $\psi(t) = \gamma e^{-\gamma t}$, so the \(n\)-fold convolution is the Erlang distribution \cite{feller1968introduction} \(\psi^{*n}(t) = \gamma^n t^{n-1} e^{-\gamma t}/(n-1)!\)
and the probability mass function for the FPT of the discrete random walk is \cite{feller1968introduction} \(\mathbb{P}\{\tau_m = n \mid q\}= \frac{m}{n} \binom{n}{(n+m)/2} q^{(n+m)/2} (1-q)^{(n-m)/2}\).

We then find the ensemble average of the FPT density in \eqref{eq: FPT general formula} by averaging the discrete PMF, and Erlang distribution, over \(q\) and \(\gamma\), respectively:
\begin{equation} \label{FPT}
     \bar{f}_{m}(t) = \mathbb{E}_{q,\gamma}[f_{m}(t \mid q, \gamma)] = \sum^\infty_{n=m} \phi(m,n) \eta(n,t),
\end{equation}
where \(n+m\) is even and
\begin{equation}
    \phi(m,n) = \frac{m}{n} \binom{n}{\frac{1}{2}(n+m)} \frac{B(\alpha_+ + \frac{1}{2}(n+m),\alpha_-+\frac{1}{2}(n-m))}{B(\alpha_+,\alpha_-)}
\end{equation}
is the ensemble-averaged discrete FPT distribution, and
\begin{equation}
    \eta(n,t) = \frac{\beta^\alpha \Gamma(\alpha+n)}{(n-1)! \Gamma(\alpha)} \frac{t^{n-1}}{(\beta + t)^{\alpha+n}}.
\end{equation}
is the rate-averaged Erlang distribution. 

An interesting feature of the FPT density given by \eqref{FPT} is that it does not necessarily integrate to unity; particles with a net leftward bias may never reach the target boundary at \(k=m\). For an individual walker with right-jump probability \(q\), the probability of a particle arriving at the boundary in finite time is \(P_m=\mathbb{P}\{T_m < \infty \mid q\} = \min[1,(q/(1-q))^m]\) \cite{feller1968introduction,CoxMiller}, which is less than 1 whenever \(q<1/2\). Since the population distribution \(\xi(q)\) assigns nonzero probability to \(q<1/2\), a nonzero fraction of the heterogeneous population never reaches the target, and the ensemble average probability is \(\bar{P}_m = \mathbb{E}_{q}[\mathbb{P}\{T_m < \infty \mid q\}] =  \int_0^\infty \bar{f}_m(t) dt <1.\) It follows from \eqref{FPT} that \(\int_0^\infty \bar{f}_m(t) dt = \sum_{n=m}^\infty \phi(m,n)\). 
The mean FPT is infinite, as the subpopulation with \(q<1/2\) never reaches the boundary, and so contributes infinite arrival times with nonzero probability \cite{CoxMiller}. We therefore characterize typical arrival times by the mode \(t^*\), found by setting \(\sum_{n=m}^\infty \phi(m,n) \frac{\partial\eta(n,t^*)}{\partial t} = 0\), which remains well defined regardless of bias.

Figure \ref{fig: Conditional FPT} illustrates the drastic impact of heterogeneity on the first passage time density and the substantial difference between a heterogeneous model and a homogeneous model. We define the latter in which each particle takes fixed parameter values according to the means of the distributions \(\xi(q)\) and \(\zeta(\gamma)\), that is, \(\bar{q}=\alpha_+/\alpha\) and \(\bar{\gamma}=\alpha/\beta\), giving \(f_m^{\mathrm{hom}}(t)=f_m(t \mid \bar{q} ,\bar{\gamma})\).
The density for the heterogeneous population exhibits a sharp peak at short times compared to the homogeneous case and a heavier algebraic tail than the homogeneous baseline, which produces a broader distribution peaked at later times. The mode FPT is \(t_\mathrm{het}^*\approx 17.6101\) for the heterogeneous population compared to \(t_\mathrm{hom}^*\approx 208.3863\) for the homogeneous baseline, demonstrating that population heterogeneity alone dramatically enhances transport to a fixed target. 

\begin{figure}
    \centering
    \includegraphics[width=\linewidth]{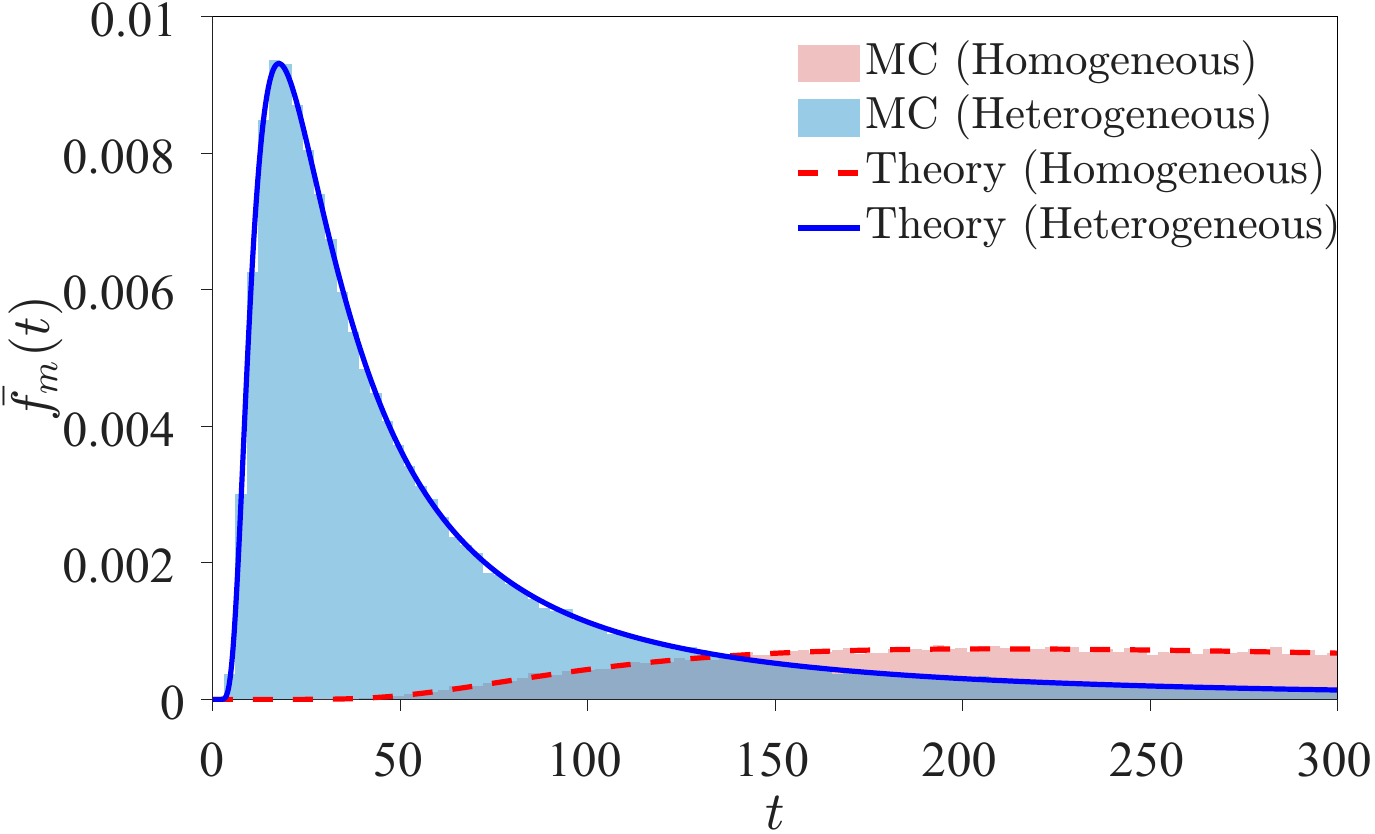}
    \caption{FPT density \(\bar{f}_m(t)\) for the heterogeneous (solid blue) and homogeneous (dashed red) populations with \(\alpha_+=\alpha_-=2\), \(\beta=1\), and \(m=50\), computed from \eqref{FPT}; for the homogeneous case, \(\bar{f}_m\) is replaced by \(f_m(t\mid \bar{q},\bar{\gamma})\). The lines show the analytical results and the histogram shows Monte Carlo simulations.}
    \label{fig: Conditional FPT}
\end{figure}

Now, we explore the impact of heterogeneity on extreme statistics. In many biological contexts, the response time is set by the first among \(N\) particles to arrive \cite{reynaud2015so,hollenbeck1989distribution,klumpp2005cooperative,reck2018cytoplasmic,higley2012calcium,basnayake2019fast,meerson2015mortality,schuss2019redundancy,
resnick1987extreme,holcman2014narrow}. Therefore, we seek to analyze the minimum FPT, \(\tau^1 = \min (t_1, \ldots, t_N)\), where \(t_i\) denotes the arrival time of the \(i\)-th particle from the group of size \(N\), and evaluate the effect that population heterogeneity has on this quantity when compared to the homogeneous baseline. For fixed values $q_i$ and $\gamma_i$, the arrival times \(t_1, t_2, \dots, t_N\) are independent. Thus, the survival probability conditional on the vectors \( \mathbf{q} = (q_1,\dots,q_N)\) and \( \boldsymbol{\gamma} = (\gamma_1,\dots,\gamma_N)\) is \(\mathbb{P}(\tau^1 > t \mid \mathbf{q}, \boldsymbol{\gamma}) =\prod_{i=1}^N \mathbb{P}(t_i > t \mid q_i,\gamma_i)\). Since the \(q_i\) and v\(\gamma_i\) are independent, we evaluate the unconditional survival probability by averaging over the gamma and beta distributions and interchanging the product and expectation, finding \(\mathbb{P}\{\tau^1>t\} = \left( \mathbb{E}_{q,\gamma} \left[  \mathbb{P}(t_i > t \mid q,\gamma) \right]
 \right)^N= [S_m(t)]^N\), where the single particle survival probability is
\begin{equation} \label{eq: survival definition}
    S_m(t)\! =\! 1\! -\! \int_0^t\! \bar{f}_{m}(s) ds\! =\! 1\! -\! \sum^\infty_{n=m}\! \phi(m,n) I\left(\frac{t}{\beta+t};n,\alpha\right),
\end{equation}
with \(I(x;a,b)\) denoting the regularized incomplete beta function (see Appendix). Then, the corresponding minimum FPT density \(f_\mathrm{min}\) defined as \cite{schuss2019redundancy}
\begin{equation} \label{f_min}
    f_{\min}(t) = -\frac{\partial}{\partial t} \left([S_m(t)]^N\right) = N[S_m(t)]^{N-1}\bar{f}_{m}(t).
\end{equation}
The equivalent formulation for the homogeneous minimum FPT density is found by replacing \(\bar{f}_m\) in equations \eqref{eq: survival definition} and \eqref{f_min} with \(f_m(t\mid \bar{q}, \bar{\gamma})\).

Figure \ref{fig: fMin Comparison With Inset} highlights an even greater difference between the minimum FPT density for the heterogeneous and homogeneous models. We consider a population size of \(N=50\) particles, with the boundary again positioned at \(m=50\).
The Monte Carlo simulations are in excellent agreement with the analytical expression \eqref{f_min}. The contrast between the two populations is clear: the heterogeneous population achieves a modal shortest arrival time of \(t_\mathrm{het}^*\approx9.1115\), compared to \(t_\mathrm{hom}^*\approx93.825\) for the homogeneous baseline, showing a tenfold reduction arising due to heterogeneity. The inset confirms that this reduction in modal arrival time persists across all \(N\).

In contrast to the mean FPT, which is infinite due to the subpopulation with \(q<1/2\) that never reaches the target, the mean shortest arrival time \(\bar{\tau}^1= \int_0^\infty [S_m(t)]^N dt\) remains effectively finite. Although \(S_m(t)\) plateaus at \(S_m(\infty)=1-\bar{P}_m\), so that \(\bar{\tau}^1\) formally diverges, the probability that all \(N\) particles fail to reach the target is \((1-\bar{P}_m)^N\), which decays exponentially with \(N\) rather than remaining fixed as for a single particle. For all \(N\) considered here, this probability is negligible, so \([S_m(t)]^N\) decays to zero well before its asymptotic plateau, and \(\bar{\tau}^1\) is well approximated by truncating the integral.

\begin{figure}
    \centering
    \includegraphics[width=\linewidth]{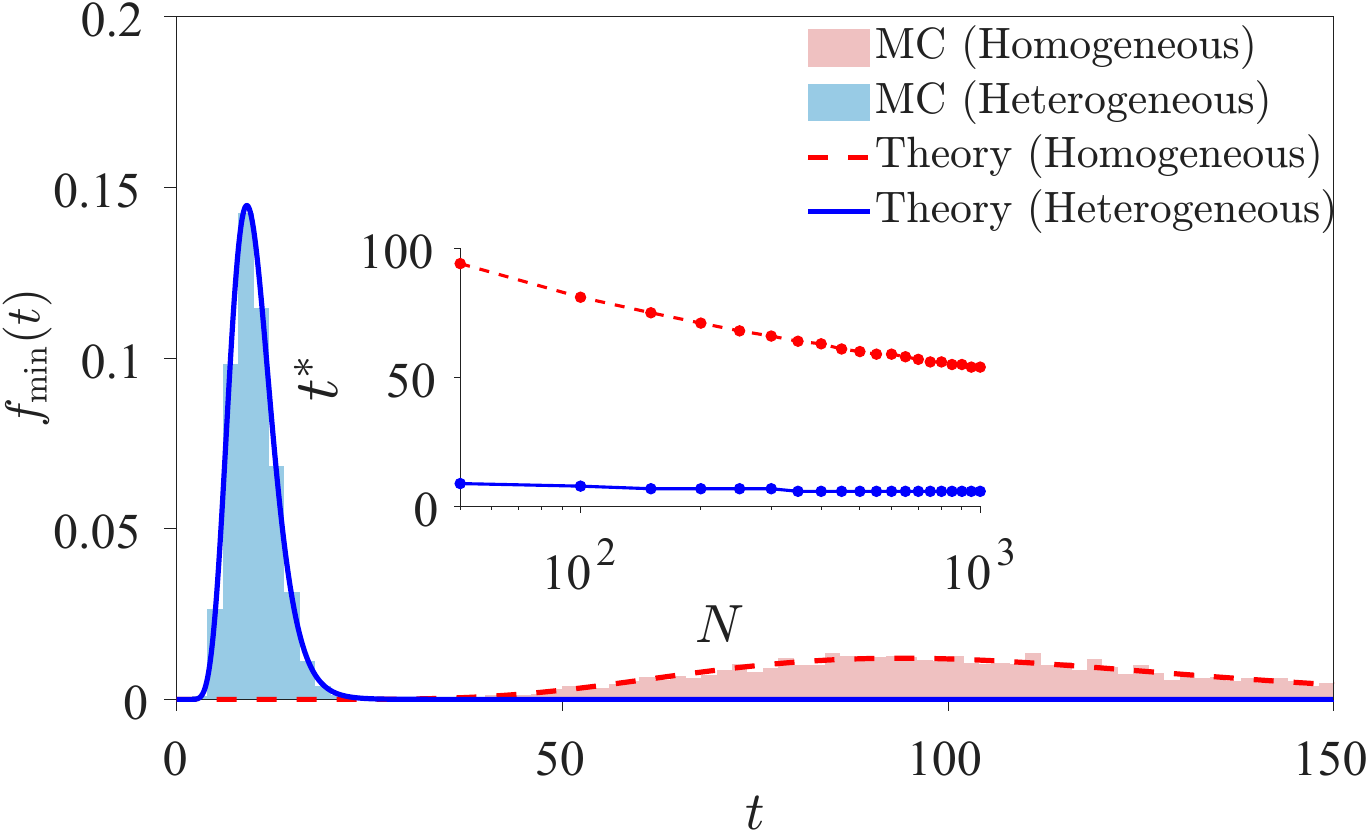}
    \caption{Minimum FPT density \(f_\mathrm{min}(t)\) for the heterogeneous (solid blue) and homogeneous (dashed red) populations of \(N=50\) particles with \(\alpha_+=\alpha_-=2\), \(\beta=1\) and \(m=50\), computed from \eqref{eq: survival definition} and \eqref{f_min}; for the homogeneous case, \(\bar{f}_m\) is replaced with \(f_m(t\mid\bar{q},\bar{\gamma})\). The lines show the analytical results and the histogram shows Monte Carlo simulations. Inset: mode shortest arrival time \(t^*\) as a function of population size \(N\) for both populations.}
    \label{fig: fMin Comparison With Inset}
\end{figure}

Now, we look at large \(N\) asymptotics, motivated by biologically relevant population sizes, such as \(\sim10^7\) spermatozoa \cite{reynaud2015so}, or \(\sim10^6\) motor proteins \cite{hollenbeck1989distribution,klumpp2005cooperative,reck2018cytoplasmic}. For \(N\gg1\), the distribution of the minimum arrival time, \(\tau^1\), is determined by the small-time behavior of the survival function \(S_m(t)\) \cite{schuss2019redundancy}. In this regime, the expansion is dominated by the \(n=m\) term, leading to the algebraic form \(1-S_m(t) \approx Ct^m\) for \(t\ll1\), where \(C\) is a scaling constant whose specific form depends on whether the population is homogeneous or heterogeneous. For the heterogeneous ensemble, the scaling constant is \(C_\mathrm{het} = B(\alpha_++m,\alpha_-)/[m\beta^mB(m,\alpha)B(\alpha_+,\alpha_-)]\). Conversely, the homogeneous baseline yields \(C_\mathrm{hom}=(\bar{q}\bar{\gamma})^m/m!\). These expansions lead to universal scaling laws for the mean and mode shortest arrival times: the mean shortest arrival time is \(\bar{\tau}^1 \approx \Gamma\left(1 + 1/m\right) (NC)^{-1/m}\), whilst the mode shortest arrival time is \(t^* \approx \left[(m-1)/{mNC}\right]^{1/m}\). 

The increased search efficiency of the heterogeneous population can be quantified by examining the ratio of the modal shortest arrival times for each population, that is, \(t^*_\mathrm{het}/t^*_\mathrm{hom} = (C_\mathrm{het}/C_\mathrm{hom})^{-1/m}\). We note that the equivalent result is obtained for the ratio of the heterogeneous and homogeneous asymptotic means. While both populations exhibit the same \(N^{-1/m}\) scaling, heterogeneity significantly reduces the arrival time through the coefficient \(C\), reflecting the dominance of a highly-biased subpopulation.

This \(N^{-1/m}\) scaling appears due to the discrete nature of the lattice, where the boundary cannot be reached in fewer than \(m\) steps. The logarithmic \(1/\ln N\) scaling, typically reported for diffusion models, emerges due to the continuum approximation utilized in such cases \cite{schuss2019redundancy,basnayake2019asymptotic}.

The parameters \(\alpha_+,\alpha_-\), and \(\beta\) carry direct physical meaning that can, in principle, be examined experimentally: \(\beta\) has units of time and sets the population's overall timescale via the mean jump rate \(\bar{\gamma}=\alpha/\beta\), while \(\alpha_\pm\) are dimensionless and control the spread and bias of the distribution independently of this timescale. In intracellular transport, \(\gamma\) is the effective motor-protein stepping rate and so relates directly to organelle velocity; the velocity distributions reported for dense-core vesicles \cite{zahn2004dense,kwinter2009dynactin} could be used to estimate \(\alpha_\pm\) and \(\beta\) for a given system. The reductions in arrival time shown in Figs. \ref{fig: Conditional FPT} and \ref{fig: fMin Comparison With Inset} are robust to such variation. Repeating the calculation of \(f_\mathrm{min}(t)\) at \(\beta=0.1\), with \(\alpha_+=\alpha_-=2,m=50,N=50\) fixed, shifts the modes to \(t_\mathrm{het}^* \approx 0.91014\) and \(t_\mathrm{hom}^* \approx 9.3615\), compared with \(t_\mathrm{het}^* \approx 9.1115\) and \(t_\mathrm{hom}^* \approx 93.825\) at \(\beta=1\). Notably, the ratio \(t^*_\mathrm{het}/t^*_\mathrm{hom}\) is unchanged, since both \(C_\mathrm{het}\) and \(C_\mathrm{hom}\) scale identically as \(\beta^{-m}\). Thus \(\beta\) sets only the speed of the transport system, while \(\alpha_\pm\) determine whether heterogeneity confers a meaningful advantage.

Throughout this paper, we have shown that population heterogeneity is not a simple addition to standard random walk theory, but rather fundamentally alters the extreme statistics that biological signaling relies upon. We have derived exact, closed-form expressions for the ensemble propagator, first-passage time density, survival function, and extreme arrival statistics of a heterogeneous population of biased continuous-time random walkers with gamma-distributed rates and beta-distributed biases. Despite sharing the same \(N^{-1/m}\) scaling law, the heterogeneous population achieves faster extreme arrival times than the homogeneous baseline with identical mean parameters, driven by a subpopulation of strongly biased, fast moving particles that dominate the tail of the FPT distribution. Our results suggest that biological systems may exploit population heterogeneity not despite the apparent inefficiency, with some particles never reaching the target, but precisely because of it: the spread of parameters guarantees the presence of a fast subpopulation whose extreme statistics are more favorable than any homogeneous ensemble with the same mean rates. 

The model is deliberately formulated in one spatial dimension on a semi-infinite lattice to isolate the effects of population heterogeneity from environmental complexity, and allow for exact analytical evaluation. Real intracellular environments introduce additional structures such as bounded domains and cytoskeletal architecture, that are not captured here, so extending the current framework to finite domains via a reflecting boundary at a second site would yield steady-state analysis, applicable to transport processes confined between two fixed cellular locations, such as repeated cargo delivery along a finite stretch of microtubule. A further modeling assumption is that each walker's parameters are drawn once at initialization and held fixed; in practice, motor protein recruitment is a dynamic process, and allowing \(\gamma\) and \(q\) to evolve stochastically over time would bridge the model with diffusing diffusivity approaches \cite{sposini2024being} and test the robustness of the self-reinforcement mechanism under temporal fluctuations.
%Finally, the focus of extreme statistics and first passage times considered here have been based solely on the shortest arrival of a single particle, and could be generalized to a \(k\)-th order threshold, where a response is only triggered when \(k\) particles have arrived. This is biologically relevant for calcium-induced calcium release \cite{higley2012calcium,basnayake2019fast}, and the structure of the \(k\)-th order extreme statistics is a natural extension to the current framework.

\bibliography{main_bibrefs}

\appendix

\maketitle

\setcounter{equation}{0}
\renewcommand{\theequation}{A\arabic{equation}}
\setcounter{figure}{0}
\renewcommand{\thefigure}{A\arabic{figure}}

\subsection*{Spatial self-reinforcement}
For fixed \(q\), the discrete master equation is
\begin{equation}
    P(k,n+1\mid q) = qP(k-1,n\mid q) + (1-q)P(k+1,n\mid q).
\end{equation}
The ensemble-averaged distribution \(\bar{P}(k,n)\) is obtained by integrating the conditional probability over \(\xi(q)\). Averaging the master equation at step \(n+1\) gives
\begin{align}
    \bar{P}(k,n+1) =\,& \int_0^1\big[qP(k-1,n\mid q) \notag\\
    &+(1-q)P(k+1,n\mid q)\big]\,\xi(q)\,dq.
\end{align}
Substituting the discrete walk solution
\(P(k,n\mid q)=\binom{n}{(n+k)/2}q^{(n+k)/2}(1-q)^{(n-k)/2}\) \cite{feller1968introduction}, the integral
\(\int_0^1 qP(k,n\mid q)\xi(q)\,dq\) reduces to a Beta function via
\(B(a+1,b)=\frac{a}{a+b}B(a,b)\) with \(a=\alpha_++\tfrac{n+k}{2}\),
\(b=\alpha_-+\tfrac{n-k}{2}\), \(a+b=\alpha+n\), giving
\begin{equation}
    \int_0^1 qP(k,n\mid q)\xi(q)\,dq
    = \left(\frac{\alpha_++\frac{n+k}{2}}{\alpha+n}\right)\bar{P}(k,n).
\end{equation}
By the symmetric identity \(B(a,b+1)=\frac{b}{a+b}B(a,b)\),
\begin{equation}
    \int_0^1(1-q)P(k,n\mid q)\xi(q)\,dq
    = \left(\frac{\alpha_-+\frac{n-k}{2}}{\alpha+n}\right)\bar{P}(k,n).
\end{equation}
Evaluating these at \(k-1\), for the \(q\)-term, and \(k+1\), for the \(1-q\)-term and substituting into the averaged master equation gives
\begin{align}
    \bar{P}(k,n+1) =\,& \frac{2\alpha_++n+k-1}{2(\alpha+n)}\,\bar{P}(k-1,n) \notag\\
    &+ \frac{2\alpha_-+n-k-1}{2(\alpha+n)}\,\bar{P}(k+1,n).
\end{align}
Writing \(\bar{P}(k,n+1)=u_n^+(k-1)\bar{P}(k-1,n)+u_n^-(k+1)\bar{P}(k+1,n)\) identifies
\begin{equation}
    u_n^\pm(k) = \frac{1}{2}\pm\frac{1}{2}\left(\frac{\alpha_+-\alpha_-+k}{\alpha+n}\right),
\end{equation}
which increases (respectively decreases) with \(k\), in other words, the spatial self-reinforcement reported in the main text.

\subsection*{Temporal self-reinforcement}
For fixed \(\gamma\), the number of jumps by time \(t\) is Poisson distributed,  \(Q(n,t\mid\gamma)=(\gamma t)^ne^{-\gamma t}/n!\). Averaging over \(\zeta(\gamma)\),
\begin{align}
    Q(n,t) &= \int_0^\infty \frac{(\gamma t)^n e^{-\gamma t}}{n!}\,\frac{\beta^\alpha\gamma^{\alpha-1}e^{-\beta\gamma}}{\Gamma(\alpha)}\,d\gamma, \notag\\
    &= \frac{\beta^\alpha t^n}{n!\,\Gamma(\alpha)}\int_0^\infty \gamma^{n+\alpha-1}e^{-(\beta+t)\gamma}\,d\gamma, \notag\\
    &= \frac{\beta^\alpha t^n}{n!\,\Gamma(\alpha)} \frac{\Gamma(n+\alpha)}{(\beta+t)^{n+\alpha}},
\end{align}
using \(\int_0^\infty x^{c-1} e^{-\lambda x} dx = \Gamma(c)/\lambda^c\). Thus
\begin{equation} \label{eq: Q(n,t)}
    Q(n,t) = \frac{\beta^\alpha\Gamma(\alpha+n)}{n!\,\Gamma(\alpha)}\, \frac{t^n}{(\beta+t)^{\alpha+n}}.
\end{equation}
Differentiating and simplifying \(n\beta - \alpha t= n(\beta +t) - (\alpha+n)t\),
\begin{align}
    \frac{d}{dt}Q(n,t)\!=\!\frac{\beta^\alpha\Gamma(\alpha+n)}{n!\,\Gamma(\alpha)} \!&\left[ \frac{n\,t^{n-1}}{(\beta+t)^{\alpha+n}}\! \right. \notag \\
    &\hspace{5mm}\left. -\frac{(\alpha+n)t^n}{(\beta+t)^{\alpha+n+1}}\!\right]\!.
\end{align}
The second term is \(\lambda_n(t) Q(n,t)\) with \(\lambda_n(t) = (\alpha+n)/(\beta+t)\). For the first term, applying \(\Gamma(\alpha+n)=(\alpha+n-1)\Gamma(\alpha+n-1)\) gives \(\lambda_{n-1}(t)Q(n-1,t)\) with \(\lambda_{n-1}(t)=(\alpha+n-1)/(\beta+t)\). Thus \(Q(n,t)\) satisfies the Kolmogrov forward equation
\begin{equation}
    \frac{d}{dt}Q(n,t)=\lambda_{n-1}(t)Q(n-1,t)-\lambda_n(t)Q(n,t),
\end{equation}
identifying \(N(t)\) as a P\'olya process \cite{CoxMiller}: \(\lambda_n(t)\) increases with \(n\) and decreases with \(t\). 

\subsection*{Ensemble Variance}
For fixed \(q\), the discrete walk \(X_n\) (step size \(a=1\)) has generating
function \(G(z)=(qz+(1-q)z^{-1})^n\) \cite{CoxMiller}. Differentiating, 
\begin{equation}
    G'(z) = n\big(qz+(1-q)z^{-1}\big)^{n-1}\big(q-(1-q)z^{-2}\big),
\end{equation}
so \(G'(z)n(2q-1)\), giving \(\mathbb{E}(X_n\mid q)=n(2q-1)\). Differentiating again,
\begin{align}
    G''(z) = &n(n-1)\big(qz+(1-q)z^{-1}\big)^{n-2}\big(q-(1-q)z^{-2}\big)^2 \notag \\
             &+ 2n(1-q)\big(qz+(1-q)z^{-1}\big)^{n-1}z^{-3},
\end{align}
so \(G''(1)=n(n-1)(2q-1)^n + 2n(1-q)\). Using \(\mathbb{E}(X_n^2\mid q)=G''(1)+G'(1)\) and \(\mathrm{Var}(X_n\mid q) =\mathbb{E}(X_n^2\mid q)-\mathbb{E}(X_n\mid q)^2\), the \(n^2 (2q-1)^2\) terms cancel, leaving
\begin{equation}
    \mathrm{Var}(X_n\mid q)=n[1-(2q-1)^2].
\end{equation}
Subordinating via \(X(t)=X_{N(t)}\) with \(N(t)\mid\gamma\sim\)
Poisson\((\gamma t)\), and using the Poisson identity \(\mathrm{Var}(N(t)\mid\gamma)=\mathbb{E}[N(t)\mid\gamma]=\gamma t\) (so \(\mathbb{E}[N^2(t)\mid\gamma]=(\gamma t)^2+\gamma t\)),
\begin{equation}
    \mathbb{E}(X(t)\mid q,\gamma)=\gamma t(2q-1), \quad
    \mathrm{Var}(X(t)\mid q,\gamma)=\gamma t.
\end{equation}
Using the standard Beta(\(\alpha_+,\alpha_-\)) and Gamma(\(\alpha,\beta\))
moments, \(\mathbb{E}[2q-1]=(\alpha_+-\alpha_-)/\alpha\),
\(\mathbb{E}[(2q-1)^2]=\big[(\alpha_+-\alpha_-)^2+\alpha\big]/\alpha(\alpha+1)\),
\(\mathbb{E}[\gamma]=\alpha/\beta\), \(\mathbb{E}[\gamma^2]=\alpha(\alpha+1)/\beta^2\),
averaging over \(q\) and \(\gamma\) gives
\begin{align}
    \langle\bar{X}(t)\rangle &= \frac{\alpha_+-\alpha_-}{\beta}\,t, \\
    \langle\bar{X}^2(t)\rangle &= \frac{(\alpha_+-\alpha_-)^2+\alpha}{\beta^2}\,t^2 + \frac{\alpha}{\beta}\,t.
\end{align}
Hence the ensemble variance is
\begin{equation}
    \mathrm{Var}(X(t)) = \langle\bar{X}^2(t)\rangle-\langle\bar{X}(t)\rangle^2
    = \frac{\alpha}{\beta^2}t^2 + \frac{\alpha}{\beta}t,
\end{equation}
where the quadratic \(t\) term signals ballistic superdiffusion. The same
result follows from the law of total variance,
\(\mathrm{Var}(X(t))=\mathbb{E}[\mathrm{Var}(X(t)\mid q,\gamma)]
+\mathrm{Var}(\mathbb{E}[X(t)\mid q,\gamma])\).

\subsection*{Single-particle survival function}
For fixed \((q_i,\gamma_i)\) the arrival times are independent, so \(\mathbb{P}(\tau^1>t\mid\mathbf{q},\boldsymbol\gamma) = \prod_{i=1}^N\mathbb{P}(t_i>t\mid q_i,\gamma_i)\). Averaging over the
i.i.d.\ \((q_i,\gamma_i)\) and using independence of expectation and product \cite{CoxMiller},
\begin{equation}
    \mathbb{P}(\tau^1>t) = \big(\mathbb{E}_{q,\gamma}[\mathbb{P}(t_i>t\mid q,\gamma)]\big)^N = [S_m(t)]^N,
\end{equation}
where \(S_m(t)=1-\int_0^t\bar{f}_m(s)\,ds\). Writing
\(\bar{f}_m(s)=\sum_{n=m}^\infty\phi(m,n)\eta(n,s)\), since all terms are non-negative the sum and integral can be exchanged, giving
\begin{align}
    S_m(t) =& 1-\sum_{n=m}^\infty\phi(m,n)J_n(t), \\[2ex]    J_n(t)=&\frac{\beta^\alpha\Gamma(\alpha+n)}{\Gamma(n)\Gamma(\alpha)} \int_0^t\frac{s^{n-1}}{(\beta+s)^{\alpha+n}}\,ds.
\end{align}
The substitution \(u=s/(\beta+s)\) (so \(\beta+s=\beta/(1-u)\), \(ds=\beta\,du/(1-u)^2\), upper limit \(u=t/(\beta+t)\)) gives \(s^{n-1}(\beta+s)^{-(\alpha+n)}ds=\beta^{-\alpha}u^{n-1}(1-u)^{\alpha-1}du\), so
\begin{align}
    J_n(t) =& \frac{1}{B(n,\alpha)}\int_0^{t/(\beta+t)}u^{n-1}(1-u)^{\alpha-1}\,du  \\
    =& I\!\left(\frac{t}{\beta+t};n,\alpha\right),
\end{align}
with \(I(x;a,b)=B(x;a,b)/B(a,b)\) the regularized incomplete Beta function. Thus,
\begin{equation} \label{eq: Survival Function Closed Form}
    S_m(t) = 1-\sum_{n=m}^\infty\phi(m,n)\, I\!\left(\frac{t}{\beta+t};n,\alpha\right).
\end{equation}

\end{document}